%% file: main.tex
\def\year{2023}\relax
\documentclass[letterpaper]{article} 
\usepackage{aaai23}  
\usepackage{times}  
\usepackage{helvet}  
\usepackage{courier}  
\usepackage[hyphens]{url}  
\usepackage{graphicx} 
\def\UrlFont{\rm}  
\usepackage{natbib}  
\usepackage{caption} 
\DeclareCaptionStyle{ruled}{labelfont=normalfont,labelsep=colon,strut=off} 
\input{pream_bm.tex}

\usepackage{makecell}
\usepackage{amsmath,amssymb, dsfont,bbm, multirow}
\usepackage{epsfig,latexsym,amssymb,amsthm,graphics,color}
\usepackage[ruled,vlined]{algorithm2e}
\SetKwInput{KwInput}{Input}
\SetKwInput{KwOutput}{Output}
\usepackage{graphicx,subfigure}
\usepackage{float,appendix}
\usepackage{booktabs}
\newtheorem{theorem}{Theorem}

\newtheorem{definition}{Definition}

\newtheorem{lemma}{Lemma}

\usepackage{enumitem}

\usepackage{newfloat}
\usepackage{listings}
\floatstyle{ruled}
\newfloat{listing}{tb}{lst}{}
\floatname{listing}{Listing}
\title{Adaptive Risk-Aware Bidding with Budget Constraint in Display Advertising}

\author {
    Zhimeng Jiang \textsuperscript{\rm 1},
    Kaixiong Zhou \textsuperscript{\rm 2}, 
    Mi Zhang \textsuperscript{\rm 3}, 
    Rui Chen \textsuperscript{\rm 3}, 
    Xia Hu \textsuperscript{\rm 2}, 
    Soo-Hyun Choi \textsuperscript{\rm 4}
}
\affiliations {
    \textsuperscript{\rm 1} Texas A\&M University,
    \textsuperscript{\rm 2} Rice University,
    \textsuperscript{\rm 3} Samsung Research America,
    \textsuperscript{\rm 4} Samsung Electronics\\
    zhimengj@tamu.edu, \{Kaixiong.Zhou,xia.hu\}@rice.edu, \{mi.zhang,rui.chen1,sh9.choi\}@samsung.com
}

\graphicspath{{figures/}}

\usepackage{bibentry}

\begin{document}

\maketitle

\begin{abstract}

Real-time bidding (RTB) has become a major paradigm of display advertising. Each ad impression generated from a user visit is auctioned in real time, where demand-side platform (DSP) automatically provide bid price usually relying on the ad impression value estimation and the optimal bid price determination. However, the current bid strategy overlooks
large randomness of the user behaviors (e.g., click) and the cost uncertainty caused by the auction competition. In this work, we explicitly factor in the uncertainty of estimated ad impression values and model the risk preference of a DSP under a specific state and market environment via a sequential decision process. Specifically, we propose a novel adaptive risk-aware bidding algorithm with budget constraint via reinforcement learning, which is the first to simultaneously consider estimation uncertainty and the dynamic risk tendency of a DSP. We theoretically unveil the intrinsic relation between the uncertainty and the risk tendency based on value at risk (VaR). Consequently, we propose two instantiations to model risk tendency, including an expert knowledge-based formulation embracing three essential properties and an adaptive learning method based on self-supervised reinforcement learning. We conduct extensive experiments on public datasets and show that the proposed framework outperforms state-of-the-art methods in practical settings.




\end{abstract}




\date{}

\section{Introduction}
In the past few years, real-time bidding (RTB) has quickly become a tens of billions of market in the globe \cite{googlewhite,yuan2013real}. In RTB, a demand-side platform (DSP) buys ad impressions in a programmatic manner on behalf of advertisers. The success of a DSP heavily relies on its \textit{bid optimization} capability \cite{wu2015predicting}, whose goal is to maximize the key performance indicator (KPI) agreed upon with advertisers (e.g., the total number of clicks or return on ad spend). In practice, bid optimization normally involves two steps, namely user response prediction and bid price determination \cite{wang2015real}. User response prediction is performed to estimate the true value of a potential ad impression. Taking the estimated value as an input, the bid price determination step aims to generate the optimal bid price for a bid request in a sequential decision making. 



In practice, RTB is a highly competitive and dynamic marketplace. The prerequisite for a DSP staying competitive is the capability of accurately predicting user responses \cite{chapelle2014modeling,ghosh2009adaptive}, e.g., click-through rate (CTR) or conversion rate (CVR). A large number of prediction models~\cite{zhang2016deep,shan2016deep,guo2017deepfm,xiao2017attentional,zhou2018deep,feng2019deep} have been proposed in the literature. All of them return a \textit{point} estimation for a bid request, which is then used to approximate the true value of the corresponding ad impression. Despite the substantial progress that has been made, their accuracy is still far from perfect. One major reason is due to incomplete data collection and unavoidable data noise~\cite{punjabi2018robust}, which result in inherent \textit{uncertainties} of estimated values. As indicated in~\cite{zhang2017managing}, explicitly factoring such uncertainties plays a critical role in optimizing a campaign's performance.

On the other hand, the dynamic nature of RTB requires modeling the correlations of bid requests under a given budget constraint in view of varying market competition. The latest research considers a DSP's bidding process as a sequential decision process and proposes model-based or model-free reinforcement learning based bidding strategies~\cite{cai2017real,wu2018budget}. While it lays a solid groundwork for bid optimization, these studies are based on the fundamental assumption that the estimations of ad impressions' values are accurate. Unfortunately, as explained before, this assumption can hardly hold in practice. We point out that a reasonable bid optimization solution needs to consider three inherently correlated components, including the uncertainties of estimations of ad impression values, the state of a DSP (e.g., remaining budget and future auction number), and market competition. The latter two components determine the DSP's \textit{risk tendency} (e.g., take more risk by bidding more aggressively or reduce risks by bidding more conservatively). 

Based on the above observations, we propose an adaptive risk-aware bidding algorithm via reinforcement learning, which, to the best of our knowledge, is the first work that simultaneously considers prediction uncertainty and the dynamic risk tendency of a DSP. We first theoretically unveil the intrinsic relation between prediction uncertainty and risk tendency, which helps generate a modified value of an ad impression. With this formulation of an ad impression value, it is critical to properly model a DSP's risk tendency. To this end, we propose two instantiations to model risk tendency, including an expert knowledge based formulation embracing three essential properties and an adaptive learning method based on self-supervised reinforcement learning. We summarize our key contributions as follows.
\begin{itemize}[wide=0pt, leftmargin=\dimexpr\labelwidth + 2\labelsep\relax]
    \item We present an adaptive risk-aware bidding algorithm, which, for the first time, considers both prediction uncertainty and risk tendency to optimize bidding performance. This framework is based on a new formulation of an ad impression value by revealing the intrinsic relation between prediction uncertainty and risk tendency. We theoretically prove that this formulation allows achieving the optimal bid price based on VaR analysis.
    \item We propose two ways to determine the risk tendency of a DSP. We identify three basic properties of risk tendency, which lead to an expert knowledge based instantiation. To mitigate the extensive manual tuning efforts, we also design a self-supervised reinforcement learning method to learn the risk tendency based on experience.
    \item We conduct extensive experiments on two public datasets to validate the superiority of our adaptive risk-aware bidding algorithm, and demonstrate the benefits of considering both prediction uncertainty and risk tendency.
\end{itemize}

\section{Problem Formulation}
In the RTB system, each bidder of a DSP acts on behalf of an advertiser and competes for advertisement auction every time a bid request is generated from a user visit. Given each auction opportunity, the bidder estimates the ad impression value and uncertainty, and then determines the bid price to maximize the cumulative ad impression value \footnote{In this paper, we use bidder and DSP interchangeably, and adopt CTR to estimate the ad impression value, while other metrics, such as CVR, can be adopted similarly. }. We aim to obtain optimal bidding strategy under the second-price auction, i.e., the bidder with highest bid price win the auction with second highest price payment.

\subsection{Problem Definition}
Considering budget constraint in real time bidding, we formulate the bid optimization problem as a Markov decision process (MDP) in the \textit{episode} level, where each episode consists of $T$ sequential bid auctions accompanied with a budget of $B$.  For each auction, we consider three pieces of critical information: (i) the remaining auction number $t\in\{0,\cdots,T\}$; (ii) the remaining budget $b\in\{0,\cdots,B\}$ and (iii) the mean value of predicted CTR (pCTR) $r_{mean}(\bx_t)$ and the corresponding standard deviation $r_{std}(\bx_t)$ for a bid request with feature vector $\bx_t$. Hence, the bidder's state $s$ is defined as $s \triangleq (t, b, \bx_t)$. Our target problem is that, given the remaining action number $t$, remaining budget $b$, the mean value of pCTR $r_{mean}(\bx_t)$ and corresponding standard deviation $r_{std}(\bx_t)$ for current bid request features $\bx_t$, how can we determine the optimal bid price $a(t, b, \bx_t)$ to optimal cumulative ad impression value in a sequential decision-making process?

\subsection{MDP Formulation}
Reinforcement learning can be represented by tuple $(\mathcal{S}, \mathcal{A}_{s}, \mathcal{P}^{s^{'}}_{sa},\mathcal{R}^{s^{'}}_{sa})$, where $\mathcal{S}$ denotes the state space, $\mathcal{A}_{s}$ denotes the action (i.e., bid price) space for state $s$, $\mathcal{P}^{s^{'}}_{sa}$ and $\mathcal{R}^{s^{'}}_{sa}$ represent the state transition probability and the immediate reward (i.e., pCTR) for transition from state $s$ to $s^{'}$ under action $a$. 
Note that $t=0$ and $b=0$ represent the end of an episode and the state with depleted budget, respectively. 

In the episode level bidding process, the state space $\mathcal{S}=\{0,\cdots,T\}\times\{0,\cdots,B\}\times \bX$, where $\bX$ denotes the set of bid request features.
Given state $s=(t,b,\bx_t)$, the action space $\mathcal{A}_s$ consists of all possible bid prices in set $\{0,\cdots,b\}$, since possible bid prices are constrained by the remaining budget $b$. Let $p_{\bx}(\bx_t)$ denote the probability of the bid request feature $\bx_t$ for a potential ad impression and $m(\delta|\bx_t)$ denote the probability of market price $\delta$ given feature $\bx_t$. 
Similar to \cite{zhang2014optimal}, we assume that market environment $m(\delta)\simeq m(\delta|\bx_t)$ , i.e., the market price distribution is independent of the bid request feature. Such independent distribution assumption can be justified by empirical evaluation via comparing winning bid distribution against different features in real-world iPinYou dataset \cite{zhang2014optimal}. For the state transition, if the bid price $a$ is larger than the market price $\delta$ (highest bid price among other competitors), then the bidder wins the ad auction, and state $(t,b,\bx_t)$ will transit to $(t-1,b-\delta,\bx_{t-1})$ with probability $p_{\bx}(\bx_{t-1})\sum_{\delta=0}^{a}m(\delta)$. Otherwise, if $a < \delta$, the bidder will lose the auction and transit to state $(t-1,b,\bx_{t-1})$ with probability $p_{\bx}(\bx_{t-1})\sum_{\delta=a+1}^{+\infty}m(\delta)$. The immediate reward is given by $r_{mean}(\bx_t)$ for $t$-th auction if the bidder wins the auction; otherwise it is $0$. 
Mathematically, the state transition probability and reward function are expressed as follows:
\be 
\mathcal{P}^{(t-1,b-\delta,\bx_{t-1})}_{(t,b,\bx_t), a}=\left\{\begin{array}{ll}
	p_{\bx}(\bx_{t-1})m(\delta),& \text{if\quad} \delta\leq a; \\
	p_{\bx}(\bx_{t-1})\sum_{\delta=a+1}^{+\infty}m(\delta),& \text{if\quad} \delta > a. 
    \end{array}\right. \nonumber
\ee 
\be
\mathcal{R}^{(t-1,b-\delta,\bx_{t-1})}_{(t,b,\bx_t), a}=\left\{\begin{array}{ll}
\theta(t,b,\bx_t),& \text{if\qquad} \delta\leq a; \\
0,& \text{if\qquad} \delta > a. 
\end{array}\right.\nonumber
\ee

\section{Methodology}
\begin{figure}[t]
\setlength{\abovecaptionskip}{-0.cm}
\setlength{\belowcaptionskip}{-0.cm}
\includegraphics[scale=0.75]{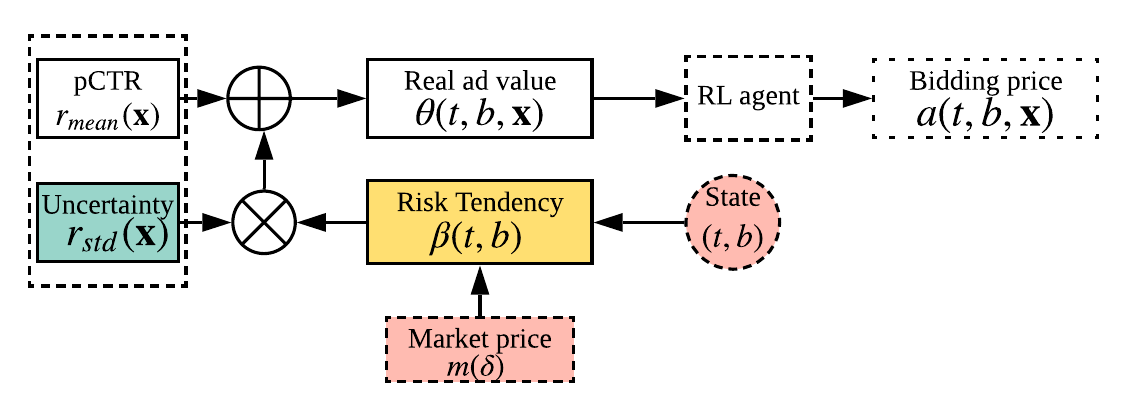}
\caption{An overview of the risk-aware reinforcement learning bidding framework. Both pCTR and prediction uncertainty come from the auction environment. Considering the market environment and bidder state, the framework combines prediction uncertainty and risk tendency to adjust ad impression values and decide bid prices.}
\label{fig.rrlb_framework}
\end{figure}

In this section, we introduce the {\bf R}isk-aware {\bf R}einforcement {\bf L}earning {\bf B}idding (RRLB) framework that effectively integrates prediction uncertainty and a bidder's risk tendency into a reinforcement learning framework. 
An overview of RRLB is illustrated in Figure~\ref{fig.rrlb_framework}. In the following sections, we first explain the uncertainties of CTR prediction with Bayesian logistic regression and then describe the risk-aware bid optimization framework and two proposed instantiations to determine the \emph{risk tendency}. Finally, we describe the model-based reinforcement learning method mapping the adjusted ad impression value to the final bid price.


\subsection{Uncertainty of CTR Prediction}\label{subsect.risk}
Similar to~\cite{zhang2017managing}, we adopt Bayesian logistic regression to explicitly measure the uncertainties of predicted CTR (pCTR) values. In Bayesian logistic regression, each weight $\bw$ is treated as a random variable instead of a parameter, and the variance of the random variable represents the uncertainty of the corresponding feature (Please see Appendix A for more details). Assuming that the weight $\bw$ follows a Gaussian distribution, Bayesian logistic regression aims to maximize the posterior weight distribution $p(\bw|\bx, y)$ given a bid request's feature vector $\bx$ and label $y$. The model output is a probability estimation of the occurrence of a click event, defined as $\hat{y}=P(y=1|\bx)$. The variance of weight is lower when the associated feature emerges more frequently, which means that such a model can measure the data completeness for each feature. Via updating the mean and covariance matrix of the weight $\bw$, the distribution of CTR $p_{\hat{y}|\bx}(\hat{y})$ can be obtained. Subsequently, we define the mean and standard deviation of CTR as $r_{mean}(\bx)=\mathbb{E}_{p_{\hat{y}|\bx}}[\hat{y}]$ and $r_{std}(\bx)=\sqrt{\mathbb{D}_{p_{\hat{y}|\bx}}[\hat{y}]}$, where $\mathbb{E}_{p_{\hat{y}|\bx}}[\cdot]$ and $\mathbb{D}_{p_{\hat{y}|\bx}}[\cdot]$ denote the expectation and variance over the CTR distribution $p_{\hat{y}|\bx}$. \emph{The standard deviation of CTR reflects the uncertainties of pCTR values}. How to obtain uncertainties of other prediction models is beyond the scope of this paper. 

\subsection{Theoretical Relation Between Uncertainty and Risk Tendency}
\label{sect.RRLB_the}

The key intuition of RRLB is to decompose the value of an ad impression $\theta(t,b,\bx_t)$ as the weighted sum of two parts: the mean pCTR and a compound term that simultaneously reflects prediction uncertainty and a bidder's risk tendency. Formally, the ad impression value is defined as follows:
\be \label{eq.modified_value}
    \theta(t,b,\bx_t)=r_{mean}(\bx_t)+\beta(t,b)r_{std}(\bx_t),
\ee
where $\beta(t,b)$ denotes the bidder's risk tendency at resource state $(t,b)$, which is a subset of bidding state $(t,b,\bx_t)$, which represents the intrinsic status of a bidder in terms of remaining auction number $t$ and remaining budget $b$ in an episode. 

Next we provide a theoretical motivation for the formulation of Eq.~(\ref{eq.modified_value}). 
The core idea is based on the value at risk (VaR) theory borrowed from finance \cite{rockafellar2000optimization,linsmeier2000value}, where VaR estimates how much the predicted CTR under/over-estimates with a given probability. Note that the goal of a bidding strategy is to improve the cumulative ad impression value. Given the current bid request feature vector $\bx_t$, remaining budget $b$ and remaining auctions number $t$, let $V^{a}(t,b,\bx_t)$ and $V^{a}_{std}(t,b,\bx_t)$ be the cumulative estimated impression value and uncertainty of the winning ads with the bidding strategy $a(t,b,\bx_t)$. Then we define VaR of the cumulative ad impression value with the bidding strategy $a(t,b,\bx_t)$ as follows:
\be
V^{a}_{\lambda}(t,b,\bx_t)\dff V^{a}(t,b,\bx_t)+\lambda(t,b) V^{a}_{std}(t,b,\bx_t),
\ee 
where state-associated coefficient $\lambda(t,b)$ is the risk preference that balances the cumulative estimated impression value and uncertainty. Note that $\beta(t,b)$ and $\lambda(t,b)$ balance the estimated value and uncertainty for the current auction and all the remaining auctions, respectively. The optimal VaR bid price $a_{VaR}$ maximizes the VaR of the cumulative ad impression value $V_{\lambda}(t,b,\bx_t)$ as follows,
\be
a_{VaR}(t,b,\bx_t)=\arg\max\limits_{0\leq a\leq b}V^{a}_{\lambda}(t,b,\bx_t).
\ee
The theorem below shows that the linear combination in Eq.~(\ref{eq.modified_value}) can achieve the optimal VaR bid price $a_{VaR}$.

\begin{theorem}[Risk Tendency Optimality]
The RRLB framework adopting the linear formulation in Eq.~(\ref{eq.modified_value}) can achieve the optimal VaR bid price $a_{VaR}(t,b,\bx_t)$ if the same risk tendency $\beta(t,b)=\lambda(t,b)$ is used.
\end{theorem}

See Appendix B for the proof. Intuitively, a rational risk tendency should be a function of the resource state, defined by $(t, b)$, of a bidder. A budget-restrained/abundant bidder would act very differently in taking risks during the bid. However, we deem that the risk tendency is independent of a random bid request $\bx$. 
\subsection{Expert Knowledge Based Risk Tendency}\label{sect.RRLB}
We leverage expert knowledge on RTB to design the first instantiation of risk tendency $\beta(t,b)$ to reveal the intrinsic risk preference of a rational bidder. 
We distill three key rules as follows. (i) The sufficiency of the remaining budget $b$ determines the sign of risk tendency $\beta(t,b)$, where a positive risk tendency indicates a strong preference to win auctions. 
(ii) The partial derivative of risk tendency $\beta(t,b)$ w.r.t. remaining budget $b$ (remaining auction number $t$) should be positive (negative) because more budget naturally allows the bidder to take the risk of bidding more ad impressions. 
(iii) When the remaining budget $b$ and remaining auction number $t$ are relatively large (e.g., the beginning of an ad campaign), risk tendency $\beta(t,b)$ depends on the ratio of $b$ to $t$ and the extent of market competition. 
Formally, we express the three rules as follows.
\begin{enumerate}[label=(\roman*), wide=0pt, leftmargin=\dimexpr\labelwidth + 4\labelsep\relax]
    \item {Sign of risk tendency:}
    \be 
    \beta(t,b) \left\{\begin{array}{ll}
	\geq 0,& b\ \mathrm{is\ sufficient\ at\ current}\ t; \\
	<0,& \mathrm{otherwise}.
    \end{array}\right.
    \ee 
    \item {Monotonicity of risk tendency:} $\frac{\partial \beta(t,b)}{\partial t}<0$, $\frac{\partial \beta(t,b)}{\partial b}>0$.
     \item {Approximation for the scenarios of large remaining budget and auction number:} $\beta(t,b)\simeq \beta(t^{'},b^{'})$ if $\frac{b}{t}=\frac{b^{'}}{t^{'}}$.
\end{enumerate}

Before elaborating the exact formulation of $\beta(t,b)$, we have to quantify the sufficiency of the budget as pointed out in the first rule. In practice, the budget richness is highly related to the level of market competition. Given a certain market price distribution $m(\delta)$ and resource state $(t,b)$, let $U(t,b)$ denote the expected bid price for an auction such that the remaining budget will be depleted in the remaining future auctions. Supposing that budget $b$ is evenly allocated to the remaining $t$ auctions, we can calculate $U(t,b)$ through the following formula:
\be
\label{eq.budget_richness}
\sum_{\delta=0}^{U(t,b)}\delta m(\delta)=\frac{b}{t}.
\ee 
Since the bidder wins an auction only if bid price $U(t,b)$ is higher than market price $\delta$, the left side of Eq.~(\ref{eq.budget_richness}) represents the expected actual cost, which should be the same as $\frac{b}{t}$ based on the assumption of even budget allocation. 

Based on the above intuition of budget richness, we formally define risk tendency as follows:
\be\label{eq.risk_alpha}
\beta(t,b)=\tanh\big(\alpha\frac{U(t,b)-\hat U}{\hat U}\big),
\ee 
where $\alpha$ is a positive hyperparameter that controls the slope of risk tendency, $\hat U$ is the budget richness threshold tuned from historical data, and function $\tanh(\cdot)$ confines risk tendency within the range $(-1,1)$. 
It can be observed that the proposed risk tendency formulation satisfies all the three expert knowledge based rules. First, the expected bid price $U(t,b)$, which measures the amount of budget richness, determines the sign of risk tendency $\beta(t,b)$, which is non-negative only if $U(t,b) \geq \hat U$ as required in rule (i). Second, both $U(t,b)$ and $\beta(t,b)$ increase with budget $b$, and decrease with $t$ as required in rule (ii). Third, the expected cost is proportional to the ratio $\frac{b}{t}$ as shown in Eq.~(\ref{eq.budget_richness}), which helps the subsequent design of risk tendency to meet rule (iii).

\subsection{Self-Supervised Risk Tendency}\label{sec:self-supervised}
\begin{figure}[t]
\setlength{\abovecaptionskip}{-0.0cm}
\setlength{\belowcaptionskip}{-0.2cm}
\includegraphics[scale=0.7]{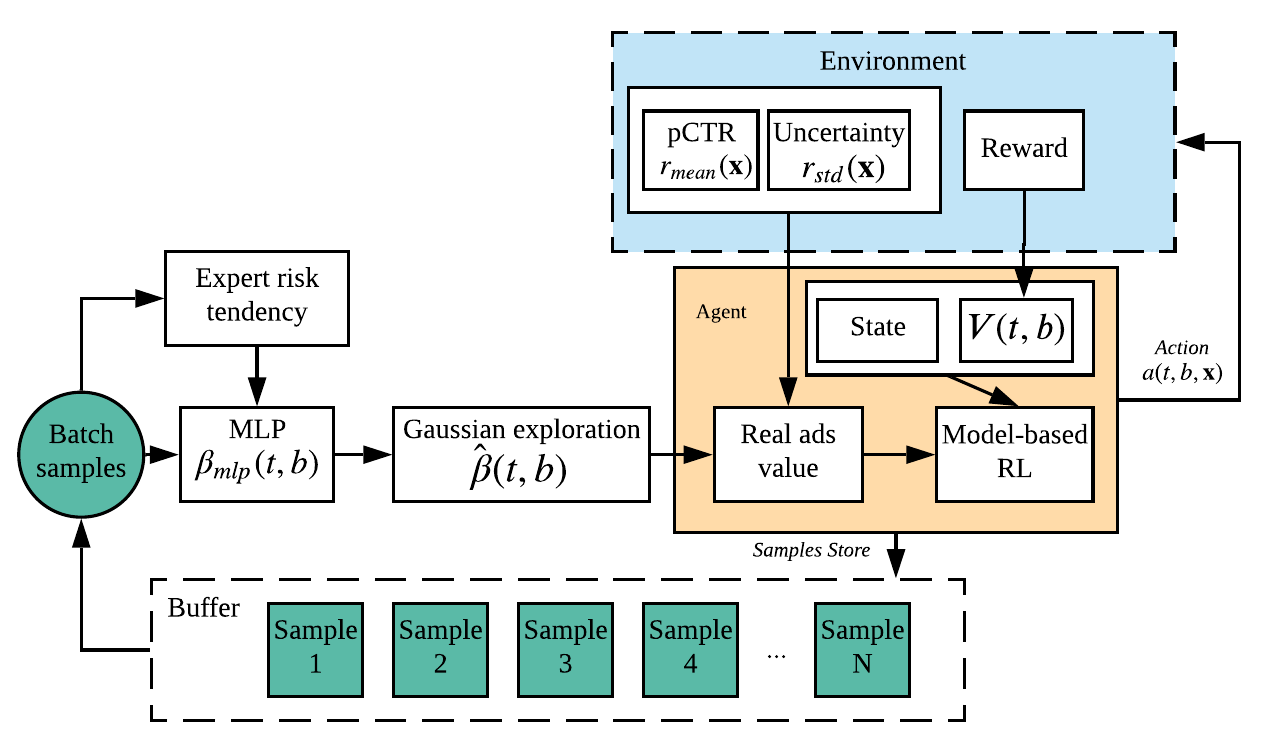}
\caption{An overview of the self-supervised reinforcement learning. The risk tendency is generated from an MLP trained by experience buffer and batch sampling in a self-supervision way. The bidder decides bid prices based on risk tendency and interacts with the RTB environment. 
}
\label{fig.ssrlb}
\end{figure}

Although the above knowledge-driven design of risk tendency may capture a bidder's intrinsic risk preference well, it requires a careful selection of hyperparameters $\alpha$ and $\hat U$ through many trials. 
To avoid such manual efforts, we propose a self-supervised reinforcement learning bidding (ssRLB) method as shown in Figure~\ref{fig.ssrlb} to automatically generate risk tendency via a multi-layer perceptron (MLP). The pseudo code of ssRLB is summarized in Appendix C.

Self supervised by the bidding history, we update the mapping function $\beta_{mlp}(t,b)=MLP(t,b;\bW_{mlp})$ generated from an MLP to approximate the risk tendency at resource state $(t,b)$ with trainable weight $\bW_{mlp}$. The ssRLB framework consists of a Gaussian exploration block, an experience buffer, an MLP mapping function, and a batch sampling. We explain the details of each component as follows.

\paragraph{Gaussian exploration.} 
We adopt exploration on risk tendency by adding Gaussian noise to $\beta_{mlp}(t,b)$ as $\hat{\beta}(t,b)= \beta_{mlp}(t,b)+\epsilon$, where noise $\epsilon\sim \mathcal{N}(0,\sigma^2)$. The noise variance $\sigma^2$ is adjustable to provide a trade-off between exploitation and exploration in reinforcement learning. 

\paragraph{Experience buffer.} Motivated by experience replay optimization in reinforcement learning \cite{zha2019experience}, we adopt the experience buffer to store good experiences represented by a quaternary set $\mathcal{B}=(t,b,\hat{\beta}(t,b),V_{episode})$ from the bidding history, where $V_{episode}$ denotes the cumulative reward for the entire episode. A ``good'' experience means that a larger reward $V_{episode}$ is obtained by using risk tendency $\hat{\beta}(t,b)$. Let $N$ be the buffer's total length. The samples with the lowest reward will be removed if the buffer is full. In this way, the experience buffer could always provide the best samples explored so far for training the MLP.

\paragraph{Batch sampling.} Batch sampling is responsible for sampling batch $\mathcal{B}_{batch}$ from the buffer. We apply a simple uniform sampling to generate $\mathcal{B}_{batch}$. 

\paragraph{Training MLP mapping function.} Given the experience batch $\mathcal{B}_{batch}$, we update the weight in the MLP mapping function by minimizing the mean square loss function:
\be \label{eq.loss}
\mathcal{L}=\sum_{(t,b,\hat{\beta}(t,b),\cdot)\in\mathcal{B}_{batch}}||MLP(t,b;\bW_{mlp}) -\hat{\beta}(t,b)||^2. \nonumber
\ee 
Since we only preserve the experiences with larger rewards in the buffer, the mapping function will be updated under supervision toward learning a good risk tendency.

\subsection{Bid Price Determination} 
\label{subsect.pricedet}
Previous sections introduce how to modify the ad impression value with prediction uncertainty and risk tendency. Next, we explain how to calculate the final bid price. We adopt the model-based reinforcement learning bidding strategy~\cite{cai2017real} to maximize the cumulative reward. Specifically, we regard the pCTR $r_{mean}(\bx_t)$ as the immediate reward for $t$-th auction, and cumulative reward $V(t,b,\bx_t)$ is defined as the expected cumulative reward starting from state $(t,b,\bx_t)$ with the optimal bid price. By definition, we have $V(0,b,\bx_t)=V(0,b)=0$ since there is no available auction. Similar to~\cite{cai2017real}, the updated policy for the cumulative reward is given by:
\be\label{eq.valueiteration0}
V(t,b,\bx_t)=\max\limits_{0\leq a \leq b}\Big\{\sum_{\delta=0}^{a}\int_{\bX}m(\delta)p_{\bx}(\bx_{t-1})\nonumber\\
\cdot\big(r_{mean}(\bx_t)+V(t-1,b-\delta,\bx_{t-1})\big)d\bx_{t-1}\nonumber\\+\sum_{\delta=a+1}^{+\infty}\int_{\bX}m(\delta)p_{\bx}(\bx_{t-1})V(t-1,b,\bx_{t-1})d\bx_{t-1}\Big\},\nonumber
\ee 
where $\bX$ represents the entire feature vector space, the two integrations represent the immediate reward for winning and losing cases. Furthermore, the cumulative reward without observation on $\bx$ can be obtained by integrating over bid request feature vector $\bx$. Formally, the cumulative reward $V(t,b)$ is:
\be \label{eq.valueiteration1}
V(t,b)\approx\max\limits_{0\leq a \leq b}\Big\{\sum_{\delta=0}^{a}m(\delta)r_{avg}+\sum_{\delta=0}^{a}m(\delta)V(t-1,b-\delta)\nonumber\\ +\sum_{\delta=a+1}^{\infty}m(\delta)V(t-1,b)\Big\}, \qquad
\qquad\qquad \nonumber
\ee 
where $r_{avg}=\int_{\bX}p_{\bx}(\bx_{t-1})r_{mean}(\bx_{t-1})d\bx_{t-1}$ is the average ad impression value over the entire feature vector space. The cumulative reward $V(t,b)$ can be iteratively updated given the average ad impression value and market price distribution. Note that $\sum_{\delta=0}^{\infty}m(\delta)=1$. The bid price at state $(t,b,\bx_t)$ is calculated by:  
\be
a(t,b,\bx_t)&\dff&\arg\max\limits_{0\leq a \leq b}V(t,b,\bx_t) \nonumber\\
&=&\arg\max\limits_{0\leq a \leq b}\Big\{\sum_{\delta=0}^{a}m(\delta)\Big(\theta(t, b, \bx_t)\nonumber\\&&+V(t-1, b-\delta)-V(t-1, b)\Big)\Big\} \nonumber\\
&=&\arg\max\limits_{0\leq a \leq b}\Big\{\sum_{\delta=0}^{a}m(\delta)g(\delta)\Big\},  \nonumber
\ee
where $g(\delta)\dff \theta(t, b, \bx_t)+V(t-1, b-\delta)-V(t-1, b)$. The cumulative reward $V(t,b)$ monotonically increases w.r.t. the remaining budget $b$, and $g(\delta)$ monotonically decreases. If $g(b)<0$, there must exist an integer price $A$ satisfying that $0\leq A\leq b$, $g(A)\geq0$ and $g(A+1)<0$.
The optimal bid price is finally given by:
\be \label{eq.action}
a(t,b,\bx_t)=\left\{\begin{array}{ll}
	b,& \text{if\quad} g(b)\geq0; \\
	A,& \text{if\quad} g(b)<0. 
    \end{array}\right.
\ee

\section{Experiments}
In this section, we conduct experiments to evaluate our RRLB framework with the two instantiations of risk tendency, namely {\bf e}xpert {\bf k}nowledge based {\bf R}einforcement {\bf L}earning {\bf B}idding (ekRLB) and {\bf s}elf-{\bf s}upervised {\bf R}einforcement {\bf L}earning {\bf B}idding ({\bf ssRLB}), and answer the following two questions:
\textbf{Q1:} How does our risk-aware bid optimization solution compare with baselines in terms of the total number of clicks?
\textbf{Q2:} How do prediction uncertainty and risk tendency affect the total number of clicks?
\textbf{Q3:} How does hyperparameters affect the long-run performanceof the risk-aware bidding strategy?


\subsection{Experimental Setting}
\paragraph{Datasets.}
Our experiments are conducted on two real-world datasets, iPinYou~\footnote{https://contest.ipinyou.com/} and YOYI~\footnote{http://apex.sjtu.edu.cn/datasets/7}. We follow the data preprocessing in \cite{zhang2017managing} to split training/test sets and obtain the estimations and uncertainties of ad impression values. The dataset description are as follows:
\begin{itemize}[wide=0pt, leftmargin=\dimexpr\labelwidth + 2\labelsep\relax]
\item {\bf{iPinYou}} dataset contains 19.5M ad impressions, 14.79K clicks and 16.0K spend (in CNY) over 9 campaigns during 10 days in year 2013. We follow the data preprocessing configuration in \cite{zhang2017managing} to split it into training/test sets, and obtain the estimations and risk of ad impression values. 
\item {\bf{YOYI}} dataset includes 402M ad impressions, 500K clicks and 428K spend (in CNY) during 8 days in year 2016. The first 7 days and the last day is set as the training data and test data, respectively.
\end{itemize}

\paragraph{Compared methods.} 
We compare ekRLB and ssRLB with two state-of-the-art baselines. Lin is a linear bidding strategy with bid price $a_{Lin}=b_0\theta(\bx)$, where parameter $b_0$ can be tuned on training data~\cite{perlich2012bid}. RLB is a model-based reinforcement learning bidding strategy~\cite{cai2017real}. These two baselines only make use of $r_{mean}(\bx)$. Besides, we also consider two variants of our proposed reinforcement learning bidding framework, one based on constant risk tendency (CRTRLB) and the other based on constant uncertainty (CURLB), for an ablation study.


\begin{table}
\setlength{\abovecaptionskip}{-0.cm}
\setlength{\belowcaptionskip}{-0.cm}
    \begin{center}
    \caption{The comparison of total click number ($c_0=1/2$).}
    \label{tab.perfcamp}
        \begin{tabular}{ c|p{0.5cm}<{\centering}p{0.5cm}<{\centering}p{0.7cm}<{\centering}p{1cm}<{\centering}p{0.9cm}<{\centering}c} 
        \toprule
        iPinYou	&Lin &RLB	&ekRLB	&CRTRLB	&CURLB & ssRLB\\ 
        \hline
        \hline
        1458	&401	&428	&428	&416	&423 & 422 \\
        2259	&19	    &73	    &75	    &59	    &70	&  70\\ 
        2261	&9	    &44	    &51	    &36	    &64	 & 61\\ 
        2821	&109	&209	&211	&198	&214 &	206\\ 
        2997	&295	&376	&382	&384	&360 &	276\\ 
        3358	&208	&233	&233	&231	&222 &	217\\ 
        3386	&157	&293	&294	&290	&293 &	290\\
        3427	&263	&290	&292	&295	&286 &	302\\ 
        3476	&206	&230	&232	&241	&233 &	204\\ \hline
        Average &185.2 & 241.8 & \textbf{244.2} & 238.9 & 240.6 & 226.9\\ \hline
        \hline
        YOYI & 725 & 890 & 894 & 873 &840 & \textbf{914} \\
        \hline
        \end{tabular}
    \end{center}
\vspace{-10pt}
\end{table}

\paragraph{Evaluation sketch.} 
Given a budget and episode length for each ad campaign, we evaluate different bidding strategies in terms of the total click number. 
In the experiments, the bidding data includes bid request feature vectors, market prices and user response (click) labels. The market adopts second-price auctions, meaning that a bidder pays only the second highest price instead of the actual bid price for a winning auction. Similar to \cite{zhang2017managing}, we calculate the pCTR $r_{mean}(\bx)$ and corresponding uncertainty $r_{std}(\bx)$ for each bid request with feature vector $\bx$. We divide training and test datasets into episodes, each of which contains $T = 1000$ auctions. As for the budget constraints, we allocate the budget as follows: $B=CPM_{train}\times10^{-3}\times T\times c_0$, where $CPM_{train}$ and $c_0$ are the cost per mille impressions in the training dataset and budget coefficient, respectively. We compare the models using the budget coefficient set $\{1/32, 1/16, 1/8, 1/4, 1/2\}$. For ekRLB, we tune the risk tendency hyperparameters $\alpha$ and $\hat U$ in Eq.~(\ref{eq.risk_alpha}) on the training dataset to optimize the number of total clicks. For ssRLB, the mapping function $MLP(t,b;\bW_{mlp})$ is realized by a four-layer MLP with 64 hidden units. Weight $\bW_{mlp}$ is trained with the Adam optimizer to minimize the mean square loss function at the learning rate of $1\times 10^{-3}$. The experience buffer size and batch size are $1\times 10^5$ and $32$, respectively. We update the buffer every $5$ training episodes. 

\subsection{Comparison Results}
We present the total click number of different methods 
on the $9$ campaigns of iPinYou and YOYI with budget coefficient $c_0=1/2$ in Table~\ref{tab.perfcamp} to answer question \textbf{Q1}. On iPinYou, ekRLB obtains the best performance on most campaigns, achieving the largest average total click number of $244.2$. On YOYI, both ekRLB and ssRLB outperform RLB, and ssRLB achieves the largest click number of 914. The two variants CRTRLB and CURLB obtain less clicks on both datasets, which validates the benefits of considering both uncertainty and risk tendency in Eq.~(\ref{eq.modified_value}). Note that ssRLB obtains less clicks than ekRLB on iPinYou but more on YOYI probably because the training of self-supervised reinforcement learning is less stable. Nevertheless, we deem that ssRLB is still a valuable alternative since it does not require manual tuning of $\alpha$ and $\hat U$ over a large volume of historical data. We leave the improvement on risk tendency learning for future work.

\begin{figure}[]%
\includegraphics[width=\linewidth]{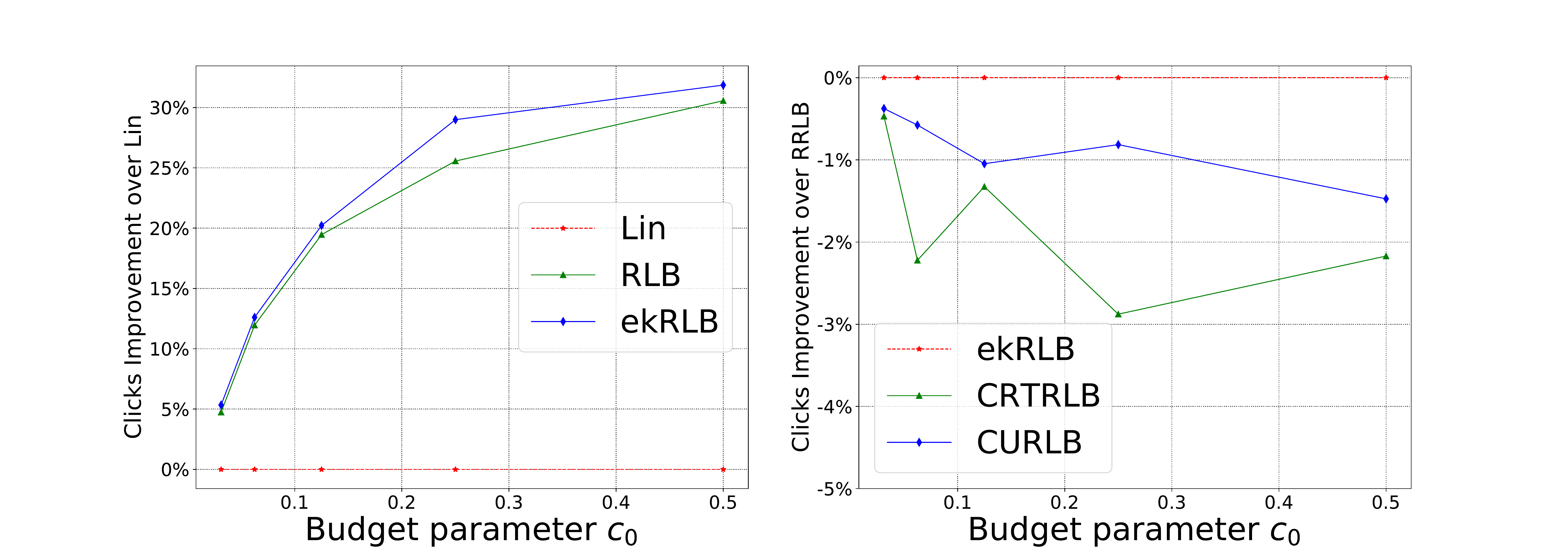}%
\caption{Left: The click number improvements of RLB and ekRLB over Lin in all budget coefficients on iPinYou. Right: The click number diminution of CRTRLB and CURLB over ekRLB on iPinYou.}\label{fig.click_no_improvement}
\vspace{-15pt}
\end{figure}

In the left part of Figure~\ref{fig.click_no_improvement}, we further show the performance improvements of RLB and ekRLB over Lin on iPinYou under the entire budget coefficient set. Two notable findings can be observed. First, the two reinforcement learning based strategies outperform Lin consistently over all budget settings, which validates the benefits of modeling the bidding process as an MDP. Given a specific budget constraint, all bid requests are inherently correlated instead of independent in Lin since previous bid prices determine the remaining budget. 
Second, ekRLB gets more clicks than the traditional RLB, especially in the cases with larger budgets, which demonstrates the advantage of explicitly modeling prediction uncertainty within a risk-aware reinforcement learning framework.

\subsection{Ablation Study}
To answer \textbf{Q2}, we study the individual contributions of prediction uncertainty and risk tendency by comparing with constant uncertainty and risk tendency. 
Regarding ekRLB as the benchmark method, the right part of Figure \ref{fig.click_no_improvement} shows the performance degradation of the two variants CRTRLB and CURLB. Along with the results in Table~\ref{tab.perfcamp}, 
we can find that (i) both CRTRLB (with constant risk tendency) and CURLB (with constant uncertainty) perform worse than ekRLB, which proves that it is critical to consider both prediction uncertainty and risk tendency to achieve an optimal bid optimization framework. (ii) CURLB achieves better performance than CRTRLB, suggesting that modeling risk tendency is even more important than the prediction uncertainty. This is because risk tendency determines how to use (e.g., add or subtract) prediction uncertainty.

\subsection{Hyperparameter Study}
To answer \textbf{Q3}, we give a comprehensive hyperparameter study to investigate how hyperparameters slope $\alpha$, constant uncertainty $r_0$ and constant risk tendency $\beta_0$ affect the performance of the methods ekRLB, CURLB and CRTRLB, respectively. Specifically, we report the experimental results on campaign $1458$ of the iPinYou dataset with budget coefficient $c_0=1/2$ in Table~\ref{tab.alpha_tune}. The performance metrics include click number and budget consumption ratio, where budget consumption ratio is defined as the cumulative cost over the overall budget. The detailed experimental results and analysis are summarized as follows.

\textit{Slope $\alpha$ for ekRLB.}
Hyperparameter $\alpha$ controls the slope of risk tendency, and then influences the bid price of ekRLB. Note that, in the case $\alpha=0$, we have $0$ risk tendency based on Eq. (6) of the main text, and ekRLB degenerates to RLB. We can clearly observe that the total click number and profit reach the highest value at a medium slope scale of $\alpha=0.1$. 
This result implies that the total click number is not sensitive w.r.t $\alpha$ and that ekRLB is robust on the hyperparameter $\alpha$. 


\textit{Constant uncertainty $r_0$ for CURLB.}
The method CURLB achieves relative comparable performance with ekRLB even though a constant uncertainty is applied. We further study the influence of constant uncertainty $r_0$ by varying it in range $[0, 1.8\times \bar{r}_{std}]$. Here we set constant uncertainty $r_0$ as a constant coefficient multiplying with the average uncertainty $\bar{r}_{std}$ in the training dataset. For the case $r_0=0$, we have prediction uncertainty of $0$ for all ad impressions, when CURLB degenerates to RLB.  We observe that constant $r_0=0.2*\bar{r}_{std}$ achieves the best performance in terms of both click number and profit, since higher/lower risks lead to ad impressions with overly large/small prices that decrease the total click number.


\textit{Constant risk tendency $\beta_0$ for CRTRLB.} Compared with ekRLB, it can been observed that a random selection of constant risk tendency $\beta_0$ in CRTRLB greatly damages model performance. We further study its influences by considering $\beta_0$ within range $[-0.5, 0]$. We only use the negative risk tendencies since a positive one tends to overestimate an ad impression and usually results in worse performance. For the case $\beta_0=0$, we remove risk tendency and degenerate CRTRLB to RLB. We observe that $\beta_0=0.0$ achieves the most clicks and that when $\beta_0$ equals $-0.4$ to $-0.5$, CRTRLB extremely underestimates the ad impression value and bids with a small price, leading to losing almost all ad impressions and low budget consumption ratio.


\begin{table}[t]
\setlength{\abovecaptionskip}{-0.cm}
\setlength{\belowcaptionskip}{-0.cm}
\setlength{\tabcolsep}{13pt}
    \begin{center}
    \caption{Hyperparameter study on slope $\alpha$ (ekRLB), uncertainty $r_0$ (CURLB) and risk tendency $\beta_0$ (CRTRLB).}
    \label{tab.alpha_tune}
        \begin{tabular}{ c|c|p{0.1cm}<{\centering} p{3cm}<{\centering} } 
        \toprule
        \multicolumn{2}{c}{Hyperparameters} & \#click & consumption ratio \\
        \hline
        \hline
        \multirow{6}*{\textbf{$\alpha$}} 
        & 0.0	&1928	&98.70\% \\
        
        & 0.001	&1925	&98.71\% \\
                
        & 0.01	&1927	&98.86\% \\
                
        & 0.1	&\textbf{1930}	&99.54\% \\
                
        & 0.2	&1923	&99.00\% \\
                
        & 0.3	&1923	&99.11\% \\
        
        & 0.4	&1924	&99.20\% \\
        
        & 0.5	&1928	&99.29\% \\
                
                
                
                
        \hline
        \multirow{5}*{\textbf{$r_0$}} 
        & 0 &	1928	&99.83\% \\ 
        & $0.2*\bar{r}_{std}$	&\textbf{1932}	&99.88\% \\ 
        & $0.4*\bar{r}_{std}$	&1922	&99.89\% \\ 
        & $0.6*\bar{r}_{std}$	&1919	&99.90\% \\ 
        & $0.8*\bar{r}_{std}$	&1910	&99.90\% \\ 
        & $1.0*\bar{r}_{std}$	&1898	&99.91\% \\ 
        & $1.2*\bar{r}_{std}$	&1896	&99.91\% \\ 
        & $1.4*\bar{r}_{std}$	&1884	&99.91\% \\ 
        & $1.6*\bar{r}_{std}$	&1878	&99.91\% \\ 
        & $1.8*\bar{r}_{std}$	&1876	&99.91\% \\
        \hline
        \multirow{6}*{\textbf{$\beta_0$}} 
        & 0.0	&\textbf{1928}	&98.69\% \\ 
        & -0.001	&1925	&98.68\% \\ 
        & -0.01	&1922	&98.55\% \\ 
        & -0.1	&1857	&96.50\% \\ 
        & -0.2	&1629	&90.00\% \\ 
        & -0.3	&845	&32.77\% \\ 
        & -0.4	&159	&5.25\% \\ 
        & -0.5	&0   	&0.06\% \\  \hline
        \end{tabular}
    \end{center}
\end{table}
\subsection{Visualization of Risk Tendency}
\begin{figure}[t]%

\centering

\subfigure{
\includegraphics[width=0.47\linewidth]{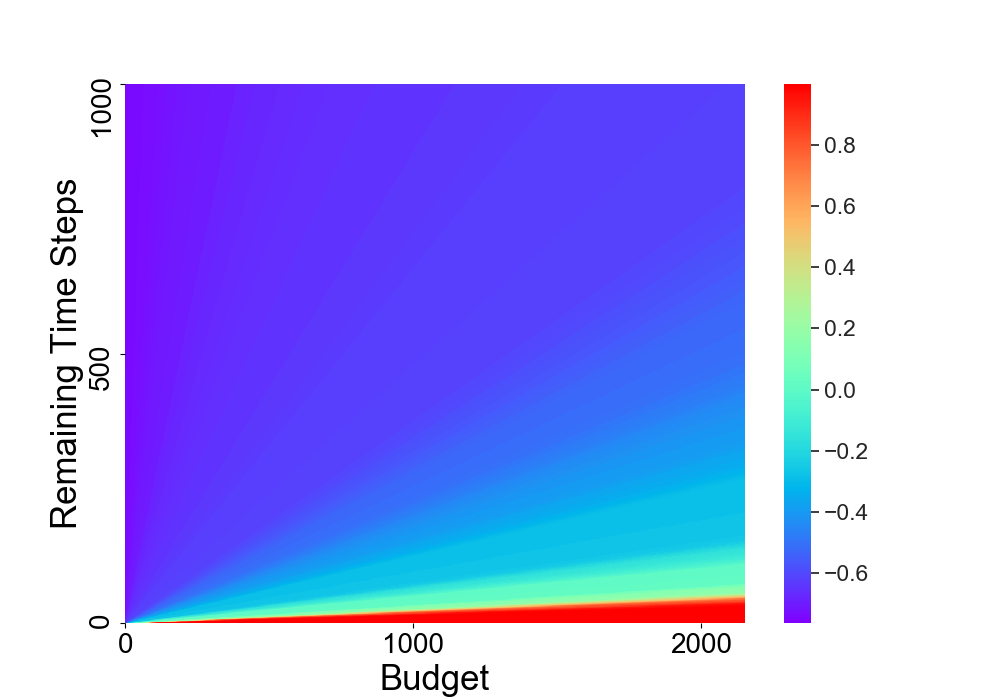}
}
\subfigure{
\includegraphics[width=0.47\linewidth]{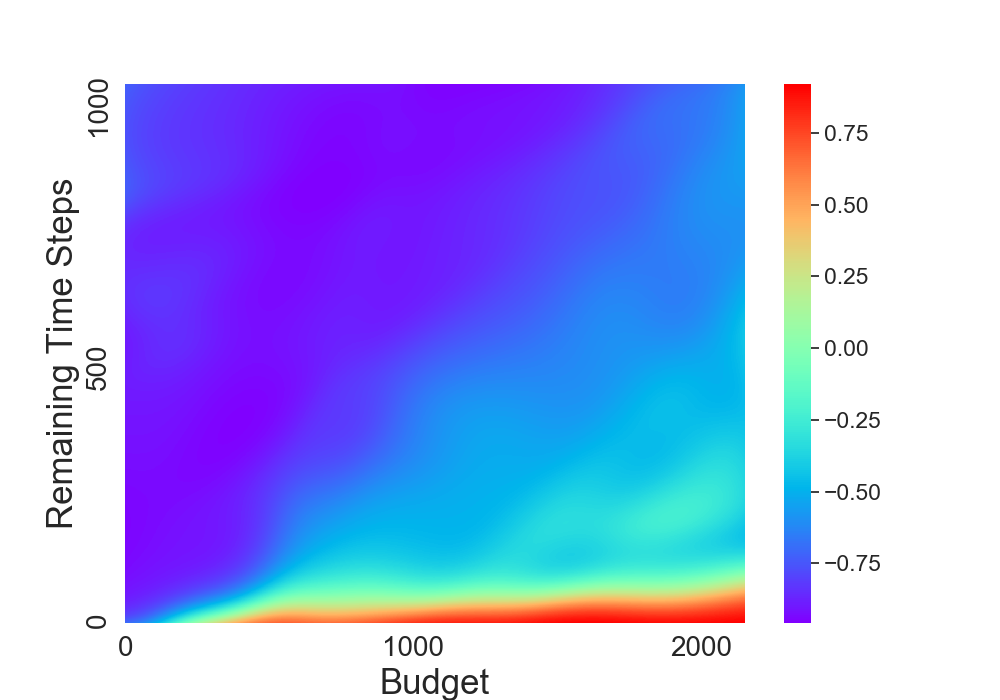}
}
\caption{Left: Risk tendency obtained by expert knowledge. Right: Risk tendency learned from self supervision. X-axis and Y-axis represent remaining budget $b$ and remaining auctions $t$, respectively.}
\label{fig.risk_tendency}
\end{figure}

Risk tendency reflects the risk preference of a rational bidder on given states. 
We designed two different methods to learn risk tendency, one based on expert knowledge and the other based on self-supervised learning. We visualize these two methods' risk tendencies for all states in Figure~\ref{fig.risk_tendency}. It can be observed that both approaches have similar trends in mapping states to risk tendency, which suggests that our self-supervised learning algorithm aligns well with the expert knowledge. Specifically, risk tendencies are negative for those resource-limited states at the upper left corner, where a bidder has a small budget for a large number of remaining auctions. In such a state, the bidder prefers to bid with conservative prices. On the other hand, risk tendencies change to positive for those resource-rich states at the bottom right corner because the bidder has a sufficient budget for remaining auctions to support more aggressive bid prices.

\section{Related Work}

{\bfseries User response prediction.} 
User response prediction can be modeled as a probability estimation task, e.g., click-through rate (CTR) \cite{graepel2010web}, conversion rate (CVR) \cite{mcmahan2013ad}. A series of prediction models have been proposed for user response prediction, including linear model~\cite{lee2012estimating}, factorization machines \cite{oentaryo2014predicting} and deep learning based models \cite{guo2017deepfm,zhou2018deep,feng2019deep}. 


\noindent{\bfseries Bidding strategy.} 
Truthful bidding is the most fundamental strategy that has been proven to be optimal in second-price auctions with unlimited budget~\cite{krishna2009auction}. In practice, every ad campaign has a budget constraint, where linear~\cite{perlich2012bid} and model-based/model-free reinforcement learning based strategies~\cite{cai2017real,wu2018budget} can be adopted to determine bid prices. \cite{zhang2017managing} is most relevant to our work, which proposes a risk management algorithm based on value at risk, but ignores intrinsic interactions between the market environment and the state of a bidder. In our work, we propose a risk-aware reinforcement learning based bidding strategy that explicitly considers such interactions.


\section{Conclusion}
In this paper, we investigated the bid optimization problem in RTB with the benefits of risk information. For the bid price determination with budget constraint, we, to the best of our knowledge, firstly consider both estimation uncertainty and the dynamic risk tendency. Specifically,
we first theoretically analyze the relation between prediction uncertainty and risk tendency of a bidder, and then proposed an adaptive risk-aware bidding algorithm with budget constraint. 
Subsequently, we developed two instantiations to determine risk tendency based on expert knowledge or self-supervised learning. Experimental results on real datasets demonstrate that RRLB making use of both prediction uncertainty and risk tendency achieves better cumulative performance than representative competitors. 



\clearpage
\bibliography{rrlb.bib}

\newpage
\appendix 
\section{Appendix A}\label{app:var}
\paragraph{VaR Introduction:} 
Our risk-aware bidding strategy adopts the linear combination $ \theta(t,b,\bx_t)=r_{mean}(\bx_t)+\beta(t,b)r_{std}(\bx_t)$ to adjust an ad impression value. In this section, we provide a theoretical motivation for such a linear combination between the pCTR and corresponding uncertainty. The core idea is based on the value at risk (VaR) borrowed from finance \cite{rockafellar2000optimization,linsmeier2000value}, where VaR estimates how much the predicted CTR under/over-estimates with a given probability. Formally, we provide a definition of VaR.

\begin{definition}
Let random variable $X$ be the profit with cumulative distribution function $F_X(x)$. The VaR at level $\alpha\in(0,1)$ is defined as the value such that the probability of a loss greater than VaR is (at most) $\alpha$ \cite{artzner1999coherent}, i.e., $VaR_{\alpha}(X)=-\inf \{x\in\mathbb{R}:F_{X}(x)>\alpha\}.$
\end{definition}

\paragraph{Probability Guarantee of Linear Combination.} Note that $V^{a}(t,b, \bx_t)$ and $V^{a}_{std}(t,b, \bx_t)$ represent the cumulative estimated ad impression value and the uncertainty of the win ads given the current bid request feature vector $\bx_t$, budget $b$, remaining impression number $t$ and the bidding strategy $a(t,b,\bx_t)$, Lemma~\ref{lemma.cantelli} states that the linear combination between $V(t,b, \bx_t)$ and $V_{std}(t,b, \bx_t)$ guarantee the probability of one-sided tail for random CTR distribution, which is consistent with VaR in finance.

\begin{lemma}[Cantelli's inequality] \label{lemma.cantelli}
For a real-valued random variable $X$ with mean $\mu$ and standard deviation $\sigma$, the following inequality holds:
\be 
\mathbb{P}(X<\mu+\lambda\sigma)<\frac{1}{1+\lambda^2}, \lambda<0, \nonumber\\
\mathbb{P}(X>\mu+\lambda\sigma)<\frac{1}{1+\lambda^2}, \lambda>0. \nonumber
\ee 
\end{lemma}
The above probability bounds motivate the definition of the VaR of cumulative ad impression value $V^{a}_{\lambda}(t,b,\bx_t)\dff V^{a}(t,b,\bx_t)+\lambda(t,b) V^{a}_{std}(t,b,\bx_t)$.

\section{Appendix B} \label{app:proof}
\paragraph{The proof of Theorem 1 (Risk Tendency Optimality)}
\begin{proof}
Note that the bidder wins an ad impression if the bid price $a$ larger than the market price $\delta$. The cumulative estimated ad impression value and the corresponding uncertainty should satisfy
\be 
V^{a}(t,b,\bx_t)&=&\sum_{\delta=0}^{a(t,b,\boldsymbol{x}_t)}m(\delta)r_{mean}(\bx_t)\nonumber\\&&+\sum_{\delta=0}^{a(t,b,\boldsymbol{x}_t)}m(\delta)V(t-1,b-\delta)\nonumber\\&&+\sum_{\delta=a(t,b,\boldsymbol{x}_t)+1}^{\infty}m(\delta)V^{a}(t-1,b),\label{eq.v1}
\ee 
\be
V^{a}_{std}(t,b,\bx_t)&=&\sum_{\delta=0}^{a(t,b,\boldsymbol{x}_t)}m(\delta)r_{std}(\bx_t)\nonumber\\&&+\sum_{\delta=0}^{a(t,b,\boldsymbol{x}_t)}m(\delta)V_{std}(t-1,b-\delta)\nonumber\\&&+\sum_{\delta=a(t,b,\boldsymbol{x}_t)+1}^{\infty}m(\delta)V_{std}(t-1,b),\label{eq.v2}
\ee
Combining Eq.~(\ref{eq.v1}) and Eq.~(\ref{eq.v2}), we have 
\be 
V^{a}_{\lambda}(t,b,\bx_t)&=&\sum_{\delta=0}^{a(t,b,\boldsymbol{x}_t)}m(\delta)\big(r_{mean}(\bx_t)+\lambda(t,b) r_{std}(\bx_t)\big)\nonumber\\&&+\sum_{\delta=0}^{a(t,b,\boldsymbol{x}_t)}m(\delta)V^{a}_{\lambda}(t-1,b-\delta)\nonumber\\&&+\sum_{\delta=a(t,b,\boldsymbol{x}_t)+1}^{\infty}m(\delta)V^{a}_{\lambda}(t-1,b),
\ee 
By firstly setting $\theta(t,b,\bx_t)=r_{mean}(\bx_t)+\beta(t,b)r_{std}(\bx_t)$ and adopting the model-based method \cite{cai2017real}, it can be seen that the linear combination in  $\theta(t,b,\bx_t)$ can achieve the optimal bid price $a_{VaR}(t,b,\bx_t)$, which maximizes the VaR of the cumulative ad impression value.
\end{proof}

\section{Appendix C} \label{app:algo}
\paragraph{Training Algorithm:} The pseudo code algorithm for self-supervised risk tendency learning is shown in Algorithm.\ref{algo.ssrlb}.

\begin{algorithm}[t]
\SetAlgoLined
\KwInput{The historical data sample with pCTR, uncertainty, market price, click labels, episode length $T$, and budget $B$}
\KwOutput{Optimal bid price}
Initialize the risk tendency, uniform replay policy \;
Update cumulative reward $V(t,b)$ \;
    \For{each episode}{
        \For{each ad auction in current episode}{
        calculate bid price based on Eq. (\ref{eq.modified_value}), execute auction and observe $(t-1, b)$ and cumulative reward starting from initial state\;
        }
        Calculate the cumulative reward of an episode\;
        \uIf{the cumulative reward $V_{episode}$ is larger than lowest cumulative reward in Buffer}{
        $\mathcal{B} \gets (t,b, \hat{\beta}(t,b), V_{episode})$\;}
        Uniformly sample a batch $\mathcal{B}_s$ from $\mathcal{B}$ \;
        Train the MLP based on the batch sample \;
        Update risk tendency $\hat{\beta}(t,b)$\ and cumulative reward $V(t,b)$;
    }
\caption{Self-supervised risk tendency learning}
\label{algo.ssrlb}
\end{algorithm}
\end{document}

%% file: pream_bm.tex
\newtheorem{prop}{Proposition}

\newtheorem{cor}{Corollary}

\newtheorem{lm}{Lemma}

\newtheorem{thm}{Theorem}

\newcommand{\be}{\begin{eqnarray}}
\newcommand{\ee}{\end{eqnarray}}
\newcommand{\benn}{\begin{eqnarray*}}
\newcommand{\eenn}{\end{eqnarray*}}
\def\IR{\rm I \kern-0.20em R}
\newcommand{\utwi}[1]{\mbox{\boldmath $ #1$}}

\newcommand{\bthm}{\begin{thm}}
\newcommand{\ethm}{\end{thm}}

\newcommand{\bcor}{\begin{cor}}
\newcommand{\ecor}{\end{cor}}
\newcommand{\bprop}{\begin{prop}}
\newcommand{\eprop}{\end{prop}}
\newcommand{\blm}{\begin{lm}}
\newcommand{\elm}{\end{lm}}
\newcommand{\beq}{\begin{equation}}
\newcommand{\eeq}{\end{equation}}
\newcommand{\ber}{\begin{eqnarray}}
\newcommand{\eer}{\end{eqnarray}}

\newcommand{\bproof}{\begin{proof}}
\newcommand{\eproof}{\end{proof}}

\newcommand{\bit}{\begin{itemize}}
\newcommand{\eit}{\end{itemize}}
\newcommand{\ben}{\begin{enumerate}}
\newcommand{\een}{\end{enumerate}}
\newcommand{\bdesc}{\begin{description}}
\newcommand{\edesc}{\end{description}}
\newcommand{\beqarrn}{\begin{eqnarray*}}
\newcommand{\eeqarrn}{\end{eqnarray*}}
\newcommand{\bproofof}{\begin{proofof}}
\newcommand{\eproofof}{\end{proofof}}
\newenvironment{rem}{\begin{trivlist}\item[]{\bf
Remark:}\hspace{4mm}}{\end{trivlist}}
\newcommand{\brem}{\begin{rem}}
\newcommand{\erem}{\end{rem}}
\newenvironment{rems}{\begin{trivlist}\item[]{\bf
Remarks}\begin{itemize}}{\end{itemize}\end{trivlist}}
\newcommand{\brems}{\begin{rems}}
\newcommand{\erems}{\end{rems}}
\newtheorem{fact}{Fact}
\newcommand{\bfact}{\begin{fact}}
\newcommand{\efact}{\end{fact}}
\newtheorem{examp}{Example}
\newcommand{\bexamp}{\begin{examp}\rm}
\newcommand{\eexamp}{\end{examp}}
\newtheorem{defn}{Definition}
\newcommand{\bdefn}{\begin{defn}\rm}
\newcommand{\edefn}{\end{defn}}

\newtheorem{alg}{Algorithm}
\newcommand{\balg}{\begin{alg}}
\newcommand{\ealg}{\end{alg}}

\newtheorem{prob}{Problem}
\newcommand{\bprob}{\begin{prob}}
\newcommand{\eprob}{\end{prob}}

\newcommand{\bvtm}{\begin{verbatim}}
\newcommand{\bfig}{\begin{figure}}
\newcommand{\efig}{\end{figure}}
\newcommand{\bcen}{\begin{center}}
\newcommand{\ecen}{\end{center}}

\long\def\comment#1{}

\def \n2{{N_0 \over 2}}

\def \h5{\hspace{0.5in}}

\newcommand{\bw}{{\utwi{w}}}
\newcommand{\bx}{{\utwi{x}}}

\newcommand{\bW}{{\utwi{W}}}
\newcommand{\bX}{{\utwi{X}}}

\newcommand{\dff}{\stackrel{\triangle}{=}}